\documentclass{svproc}
\usepackage{url}
\usepackage{graphicx}
\usepackage{amssymb}

\def\beq{\begin{equation}}
\def\eeq{\end{equation}}
\def\bea{\begin{eqnarray}}
\def\eea{\end{eqnarray}}

\def\ddt{\partial_t}
\def\ddx{\partial_x}

\def\ddr{\partial_r}

\def\dd{\mbox{$\mathrm d$}}
\def\dt{\mbox{d}t}

\def\nablav{\mbox{\boldmath$\nabla$}}
\def\grad{\nablav}
\def\rot{\nablav\times}
\def\div{\nablav\cdot}

\def\Av{{\bf A}}
\def\Ev{{\bf E}}
\def\Bv{{\bf B}}
\def\Fv{{\bf F}}
\def\Jv{{\bf J}}
\def\Lv{{\bf L}}
\def\Sv{{\bf S}}
\def\Vv{{\bf V}}
\def\av{{\bf a}}
\def\gv{{\bf g}}
\def\rv{{\bf r}}
\def\pv{{\bf p}}
\def\uv{{\bf u}}
\def\vv{{\bf v}}

\def\nv{\hat{\bf n}}
\def\xv{\hat{\bf x}}
\def\yv{\hat{\bf y}}
\def\zv{\hat{\bf z}}
\def\bra{\left\langle}

\def\ket{\right\rangle}
\newcommand{\braket}[1]{\bra#1\ket}

\begin{document}
\mainmatter

\title{Basics of Laser-Plasma Interaction:\\ a Selection of Topics}

\titlerunning{Basics of Laser-Plasma Interaction}

\author{Andrea Macchi}
\authorrunning{A. Macchi} 
\tocauthor{Andrea Macchi}
\institute{National Institute of Optics, National Research Council (CNR/INO),\\
Adriano Gozzini laboratory, via Giuseppe Moruzzi 1, 56124 Pisa, Italy,\\
\email{andrea.macchi@ino.cnr.it},\\ 
and\\
Enrico Fermi Department of Physics, University of Pisa, largo Bruno Pontecorvo 3, 56127 Pisa, Italy
}

\maketitle

\begin{abstract}
A short, tutorial introduction to some basic concepts of laser-plasma interactions at ultra-high intensities is given. The selected topics include a) elements of the relativistic dynamics of an electron in electromagnetic fields, including the ponderomotive force and classical radiation friction; b) the ``relativistic'' nonlinear optical transparency and self-focusing; c) the moving mirror concept and its application to light sail acceleration and high harmonic generation, with a note on related instabilities; d) some specific phenomena related to the absorption of energy, kinetic momentum and angular momentum from the laser light.   
\end{abstract}
\section{Introduction}
Present-day short pulse, high power laser systems have reached the petawatt ($10^{15}$~W) level. When such power is tightly focused in a spot with a diameter of few wavelengths $\lambda$ ($\simeq 1~\mu$m for sub-picosecond systems), intensities exceeding $10^{21}$~W~cm$^{-2}$ may be achieved. The corresponding strength of the EM fields is such that any sample of matter exposed to such fields becomes instantaneously highly ionized, i.e. turned into a plasma, and the freed electrons oscillate with momenta largely exceeding $m_ec$ (where $m_e$ is the electron mass and $c$ is the speed of light). The nonlinear dynamics of such relativistic plasma in a superstrong EM field is the basis of advanced schemes of laser-plasma sources of high energy electrons, ions and photons which are characterized by high brilliance and ultrashort duration.

A few years ago we tried to present the basic concepts of the theory of superintense laser-plasma interactions in a primer of about one hundred of pages \cite{macchi-book}, and it is hard to further condensate such material. Thus, the present paper is mostly an ultrashort introduction to the field at a ``sub-primer'' level, focused on an arbitrary selection of contents. We do not enter into mathematical details which can be found in the primer or in the other (few) references we cite. 

Our rough selection criterion is to include here preferentially topics on which either we witnessed frequent misunderstanding or we may add something with respect to our primer. Beyond the latter, more complete and advanced introductions may be found in textbooks \cite{gibbon-book,mulser-book} or review papers \cite{bulanovRMP06,gibbonRNC12}. We also address the reader to other reviews for the important topics of laser-plasma accelerators of both electrons \cite{esareyRMP09} and ions \cite{macchiRMP13}, on which additional references may be found in other contributions to this book. On topics where controversies are present, we have only room to give our personal point of view.

\section{Single electron dynamics and radiation friction}
\label{sec:single}

A look at the dynamics of a single electron in an EM field of arbitrary amplitude is a good warm-up before discussing a many-particle system with collective effects, i.e. a plasma. 
In non-covariant notation, the relativistic motion of an electron in a \emph{given} EM field is described by the equations
\begin{equation}
\frac{\mbox{d}\pv}{\mbox{d}t}=-e\left(\Ev+\frac\vv{c}\times\Bv\right) \; , \qquad \frac{\mbox{d}\rv}{\mbox{d}t}=\vv \; , \qquad \frac{\mbox{d}(m_e\gamma c^2)}{\mbox{d}t}=-e\vv\cdot\Ev \; ,
\label{eq:single1}
\end{equation}
where $\pv=\pv(t)$, $\rv=\rv(t)$,  $\vv={\bf v}(t)=\pv/m_e\gamma$, $\gamma=(1+\pv^2/m_e^2c^2)^{1/2}=(1-v^2/c^2)^{-1/2}$, and the fields are evaluated at the electron position, i.e. $\Ev=\Ev(\rv(t),t)$ and $\Bv=\Bv(\rv(t),t)$. By \emph{given} fields we mean that we neglect their self-consistent modification by the motion of the electron (see sec.\ref{sec:RF}).

\subsection{Motion in plane wave fields}
\label{sec:planewave}

Exact relations and solutions can be found for plane wave fields, conveniently described by the vector potential $\Av=\Av(x-ct)$ which we take to be propagating along $\xv$. The EM fields are given by $\Ev=-\partial_t\Av/c$ and $\Bv=\rot\Av=\xv\times\ddx\Av$. By separating the electron momentum in longitudinal ($p_x$) and transverse ($\pv_{\perp}$) components, it is possible to find two constants of motion:
\beq
\frac{\dd}{\dt}\left(\pv_{\perp}-\frac{e}{c}\Av\right)=0 \; , \qquad
\frac{\dd}{\dt}\left(p_x-m_e\gamma c\right)=0 \; . 
\label{eq:single2}
\eeq
The first relation is the conservation of canonical momentum related to the traslational invariance in the transverse plane ($yz$). The second arises from the properties of the EM field: if a net amount of energy ${\cal E}$ is absorbed from the field, a proportional amount of momentum ${\cal E}/c$ must be absorbed as well.\footnote{In fact, in classical electrodynamics the ratio between the amount of energy and of momentum modulus in a wavepacket is $c$, thus this relation must be conserved if the wavepacket is totally absorbed by a medium. In a quantum picture, one may think of the absorption of a given number of photons, each having energy ${\cal E}=\hbar\omega$ and momentum modulus ${\cal E}/c$.}
If an electron is initially at rest before it is reached by the wave, then $\pv_{\perp}=e\Av/c$ and $p_x=mc(\gamma-1)$ at any time. These relations also yield $p_x=e^2\Av^2/2m_ec^3$ and imply that, as the field is over ($\Av=0$), an electron initially at rest will be at rest again, i.e. no net acceleration is possible in a plane EM wave. 

\begin{figure}
\includegraphics[width=\textwidth]{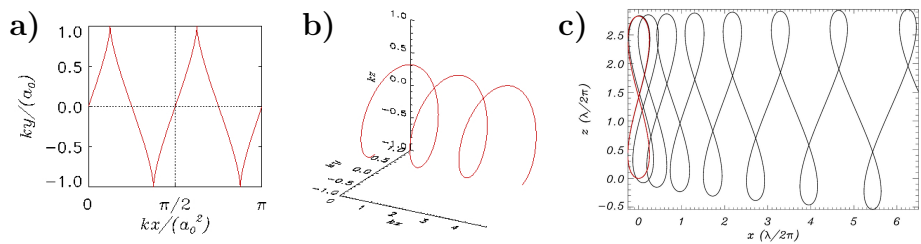}
\caption{a), b): self-similar ``drifting'' trajectories of an electron in a monochromatic plane wave for linear (a) and circular (b) polarization. c): the figure-of-eight trajectory (red line) obtained by subtracting the drift from case a), and the trajectory with same initial conditions, but adding the radiation friction force (black line).}
\label{fig:traj1}\label{fig:traj2}
\end{figure}

Now consider the case of a monochromatic wave of frequency $\omega$, 
\begin{equation}
\Av=A_0\left[\yv\cos\theta\cos(kx-\omega t)+\zv\sin\theta\sin(kx-\omega t)\right] \; , \qquad \Bv=\xv\times\Ev, 
\end{equation}
where $k=\omega/c$ and $-\pi/2<\theta<\pi/2$ determines the wave polarization: for instance $\theta=0$ and $\theta=\pm\pi/2$ correspond to linear polarization (LP), while $\theta=\pm\pi/4$ corresponds to circular polarization (CP). This wave has infinite duration, but one may still assume the same initial conditions as above if the wave is ``turned on'' over an arbitrarily long rising time. One thus obtains an average drift momentum $\braket{p_x}=\braket{e^2\Av^2/2m_ec^3}$ (the brackets denote an average over a laser period). The trajectories (Fig.\ref{fig:traj1}a-b) have a self-similar form, i.e. they can be written as function of the scaled coordinates $x/a_0^2$, $y/a_0$ and $z/a_0$ where $a_0$ is a dimensionless amplitude of the EM wave,
\beq
a_0=\frac{eA_0}{m_ec^2} \; .
\label{eq:a0}
\eeq
The drift velocity is $v_D=ca_0^2/(a_0^2+4)$. By transforming to a frame moving with such velocity along $\xv$, the trajectories become closed. For LP the electron performs a ``figure of eight'' in the plane containing $\xv$ and the polarization direction (Fig.\ref{fig:traj2}). For CP, the electron moves on a circle in the $yz$ plane. Notice that in this latter case the $\gamma$-factor is a constant and the motion does not contain high harmonics of $\omega$.

The parameter $a_0$ introduced in Eq.(\ref{eq:a0}) is a convenient indicator of the onset of the relativistic dynamics regime. In the ``no drift'' frame, the typical value of the gamma factor (temporally averaged for LP) is $\gamma=(1+a_0^2/2)^{1/2}$, thus the dynamics is strongly relativistic when $a_0\gg 1$. The parameter is related to the wave intensity $I$ and wavelength $\lambda$ by 
$a_0=0.85(I\lambda^2/10^{18}~\mbox{W cm}^{-2}\mu\mbox{m}^2)^{1/2}$.

\subsection{Ponderomotive force}
\label{sec:PF}

The motion in a plane wave is an useful reference case, but in most cases we have to deal with more complex field distributions, such as a laser pulse with a finite extension in space and time. At least we may assume the field to be
\emph{quasi-monochromatic}, i.e. to be described by  $\Av(\rv,t)=\mbox{\rm Re}\left[\tilde{\Av}(\rv,t)\mbox{\rm e}^{-i\omega t}\right]$ with $\braket{\Av(\rv,t)} \simeq 0$ and $\braket{\tilde{\Av}(\rv,t)} \simeq \tilde{\Av}(\rv,t)$, i.e. the envelope function $\tilde{\Av}(\rv,t)$ describes the temporal variation of the field on a scale slower than the oscillation at frequency $\omega$. The idea is to separate these different scales by writing for the position $\rv(t) \equiv \rv_s(t)+\rv_o(t)$ where $\braket{\rv_s(t)} \simeq \rv_s(t)$
and $\braket{\rv_o(t)} \simeq 0$, i.e. $\rv_o(t)$ describes the fast oscillation around the slowly-moving center $\rv_s(t)$. In the non-relativistic case, one obtains equations for the ``slow'' motion as
\beq
m_e\frac{\dd\vv_s}{\dt}=-\frac{e^2}{2m_e\omega^2}\grad\braket{\Ev^2(\rv_s(t),t)} \equiv \Fv_p \; , \qquad \frac{\dd\rv_s}{\dt}=\vv_s \; ,
\label{eq:pondforce}\label{eq:PF}
\eeq
where $\Fv_p$ is named the \emph{ponderomotive} force (PF).\footnote{We stress that we define the PF as a cycle-averaged approximation of the Lorentz force. However, in the literature sometimes the term ``\emph{oscillating} PF'' has been used \cite{wilksPRL92} to refer to oscillating nonlinear terms in the Lorentz force (such as the $\vv\times\Bv$ term which has a $2\omega$ component). This definition is inconsistent with the whole idea of separating the ``slow'' and ``fast'' scales in the motion.} Eq.(\ref{eq:pondforce}) is based on a perturbative approach where magnetic effects are taken into account up to first order in $v/c$, and the spatial variation of the fields over a wavelength is small ($|\lambda\grad E|\ll E$). 

\begin{figure}[t!]
\begin{center}
\includegraphics[width=0.5\textwidth]{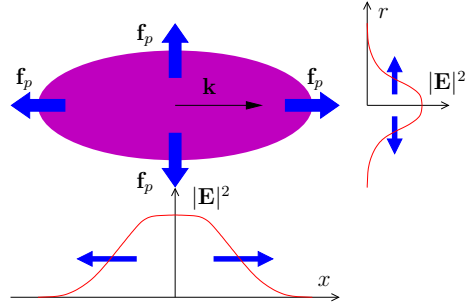}
\end{center}
\caption{Ponderomotive scattering of electrons by the ponderomotive force (\ref{eq:pondforce}) of a laser pulse having finite length and width.}
\label{fig:PFscatter}
\end{figure}

According to Eq.(\ref{eq:pondforce}) the electrons are pushed out of the regions where the field is higher. Thus, if a laser pulse propagates through a tenuous plasma (Fig.\ref{fig:PFscatter}), electrons will be pushed in the forward (propagation) direction on the leading edge of the pulse, and in the backward direction on the trailing edge: in proper conditions, this effect generates wake waves in the plasma \cite{esareyRMP09}. The PF associated to the intensity gradient in the radial direction tends to pile electrons at the edge of the laser beam and create a low-density channel along the propagation path, which can cause a self-guiding effect (see sec.\ref{sec:SF}).

An extension of the PF to the relativistic regime is not straightforward. For a quasi-transverse, quasi-plane wave field one may follow the hint that the non-relativistic PF (\ref{eq:pondforce}) is the gradient of the average oscillation energy (``ponderomotive potential''). Assuming $\pv_{\perp} \simeq e\Av/c$ and $\gamma \simeq (1+\pv_{\perp}^2/m_e^2c^2)^{1/2}$, one can write the oscillation energy in the relativistic case as $m_ec^2(\gamma-1)$ and replace the potential in (\ref{eq:pondforce}). However, one has also to take into account that the oscillatory motion yields relativistic inertia. One may thus write 
\beq
\frac{\dd}{\dt}\left(m_{\rm eff}\vv_s\right) \simeq -\grad(m_{\rm eff}c^2) \; , \qquad
m_{\rm eff} \equiv m_e(1+\braket{\av^2}(\rv_s,t))^{1/2} \; ,
\eeq
(where ${\bf a}=e{\bf A}/m_ec^2$) with $m_{\rm eff}$ acting as an effective, position- and time-dependent mass . We remark that this expression is limited to a ``semi-relativistic case'', in which the average velocity $|\vv_s|\ll c$, and for smooth field profiles where transverse components are much larger than longitudinal ones (e.g. a loosely focused laser beam).

\subsection{Radiation friction (reaction)}
\label{sec:RF}

While an electron is accelerated by an EM field, it also radiates EM waves when accelerated. But the ``standard'' equations of motion (\ref{eq:single1}) do not account for the energy and momentum carried away by the radiation. For example, according to (\ref{eq:single1}) an electron in an uniform and constant magnetic field performs a circular orbit at constant energy; but since the electron experiences a centripetal acceleration, it will radiate and lose energy, so that we expect the trajectory to become a spiral as if the electron was experiencing a {friction} force. To describe such \emph{radiation friction} (RF) effects, additional terms must be added to the Lorentz force in order that the motion is self-consistent with the radiation emission. The phenomenon can also be described as the back-action of the fields generated by the electron on itself, so it is also named \emph{radiation reaction} (RR).

RR (or RF) is a longstanding and classic problem of classical electrodynamics. In ordinary conditions the effect is either negligible or at least it can be treated perturbatively and phenomenologically, e.g. inserting a simple friction force. The dynamics of the electron becomes strongly affected by the radiation emission when the energy of the emitted radiation is comparable to the work done on the electron by the accelerating fields (\cite{jackson}, par.16.1), which implies field strengths at the frontier of those produced by present-day laser technology. This circumstance has revitalized the debate (and associated controversy) on RR in recent years. However, it is apparent that as long as a classical description is adequate, one can safely use the RR force given in the textbook by Landau and Lifshitz (LL)  \cite{landau-lifshitz-RR}: 
\beq
\Fv_{\rm RR} \simeq -\frac{2r_c^2}{3}\left(\gamma^2\left(\Lv^2-\left(\frac{\vv}{c}\cdot\Ev\right)^2\right)\frac{\vv}{c}-\Lv\times\Bv-\left(\frac{\vv}{c}\cdot\Ev\right)\cdot\Ev\right) \; ,
\label{eq:LL}
\eeq
where $\Lv\equiv \Ev+\vv\times\Bv/c$ , $r_c=e^2/m_ec^2$ is the classical electron radius, and small terms containing the temporal derivatives of the fields have been dropped down \cite{tamburiniNJP10}. It may be interesting to notice that for an electron which is instantaneously at rest ($\vv=0$) the force reduces to
\beq
\Fv_{\rm RR} \simeq \frac{2r_c^2}{3}\Ev\times\Bv=\sigma_T\frac{\Sv}{c} \; ,
\eeq
where $\sigma_T=8\pi r_c^2/3$ is the Thomson cross section for the scattering of an EM wave, and $\Sv=c\Ev\times\Bv/4\pi$ is the Poynting vector giving the energy flux of the wave (the intensity $I=|\Sv|$): thus, in this limit the RR force is a drag force which describes the absorption of an amount of EM momentum proportional to the amount of EM energy subtracted from the wave and then radiated away.

An exact solution for the motion in a plane EM wave exists also when the RR force (\ref{eq:LL}) is included \cite{dipiazzaLMP08}. The modification of the trajectory is shown in Fig.\ref{fig:traj2}, for the same initial conditions yielding the closed ``figure of eight'' when neglecting RR: if the latter is included, the trajectory opens up with the electron gaining energy and accelerating along the propagation direction. Of course a friction force sounds as unable to accelerate anything, but actually the effect of friction is to change the relative phase between the fields and the electron velocity.
This yields $\braket{\vv\cdot\Ev}\neq 0$, so that the electron gains energy from the wave, and $\braket{\vv\times\Bv}\neq 0$, so that the electron is accelerated along $\xv$.

The classical theory predicts that the spectrum of the radiation scattered from a relativistic electron peaks at frequencies $\omega_{\rm rad}\simeq \gamma^3\omega_i$ (\cite{jackson}, par.14.4), where $\omega_i$ is the frequency of the incident radiation ($\omega_{\rm rad}=\omega_i$ in the linear non-relativistic regime). Thus, with increasing $\gamma$ eventually the energy of a single photon $\hbar\omega_{\rm rad} \gtrsim m_ec^2\gamma$, the electron energy, so that the recoil from the photon emission is not negligible and a quantum electrodynamics (QED) description becomes necessary. This is reminiscent of the well-known Compton scattering, but here the relevant regime involves the sequential absorption of very many low-frequency photons and the emission of several high-frequency photons. A QED theory of RR is still an open issue and is the subject of current research (see \cite{macchiP18} for a discussion).

\section{Kinetic and fluid equations}

For a plasma of electrons and ions at high energy density, a classical approach is adequate. The most complete description of the dynamics is based on the knowledge of the distribution function $f_{a}=f_{a}({\bf r},{\bf p},t)$ which gives the density of particles in the phase space $({\bf r},{\bf p})$ for all species $a$ (e.g. $a=e,i$ for a single ion distribution). 

A great simplification arises from the possibility of neglecting binary collisions, since the cross section for Coulomb scattering quickly decreases with increasing particle energy. For further simplicity we neglect any process which may create or destroy particles (such as ionization, pair production, \ldots), as well as radiation friction (RF) whose inclusion will be discussed later. The total number of particles of each species is thus conserved, and the distribution function satisfies a continuity equation in the phase space (the Vlasov equation):
\bea
\frac{\partial f_a}{\partial t} 
   +\frac{\partial}{\partial {\bf r}}(\dot{\bf r}_af_a)
   +\frac{\partial}{\partial {\bf p}}(\dot{\bf p}_af_a)
  =0  \; ,
\label{eq:kinetic}
\eea
where
\bea
\dot{\bf r}_a={\bf v}=\frac{{\bf p}c}{({\bf p}^2+m_a^2c^2)^{1/2}} \; ,\qquad
\dot{\bf p}_a=q_a\left({\bf E}+\frac{\bf v}{c}\times{\bf B}\right) \; . 
\label{eq:kinetic2}
\eea
The coupling with Maxwell equations for the EM fields $\Ev=\Ev(\rv,t)$ and $\Bv=\Bv(\rv,t)$ occurs via the charge and current densities obtained from $f_a$:
\beq
\rho({\bf r},t)=\sum_{a} q_a \int f_a \dd^3p \; , 
\qquad 
{\bf J}({\bf r},t)=\sum_{a} q_a \int {\bf v} f_a  \dd^3p \; .
\label{eq:sources}
\eeq 
The Vlasov-Maxwell system constitutes the basis for the kinetic description of laser-plasma interactions, mostly via numerical simulations based on particle-in-cell (PIC) codes \cite{birdsall-langdon}. The PIC method may be extended to include collisions, ionization, and particle production (see e.g. \cite{arberPPCF15,derouillatCPC18}). 
RF effects can be included straightforwardly by adding the LL force \ref{eq:LL} (sec.\ref{sec:RF}) to the second of Eqs.(\ref{eq:kinetic2}).\footnote{Notice that in Eqs.(\ref{eq:kinetic}-\ref{eq:kinetic2}) $\partial_{\bf r}(\dot{\bf r}_af_a)=\dot{\bf r}_a\partial_{\bf r}f_a$ and $\partial_{\bf p}(\dot{\bf p}_af_a)=\dot{\bf p}_a\partial_{\bf p}f_a$, as it is usual to write for the Vlasov equation. However, if the LL force is added to the Lorentz force, $\partial_{\bf p}(\dot{\bf p}_af_a)\neq\dot{\bf p}_a\partial_{\bf p}f_a$. This is not an issue for the standard PIC algorithms which provide a solution of the general kinetic equation (\ref{eq:kinetic}).} The technical implementation in PIC codes proposed in Ref.\cite{tamburiniNJP10} has been successfully benchmarked in Ref.\cite{vranicCPC16}. Notice that in a simulation, because of the finite resolution of a spatial grid over which the fields are represented, it is almost impossible to resolve the high-energy radiation emitted by ultra-relativistic electrons at frequencies $\omega_{\rm rad} \simeq \gamma^3\omega$, with $\omega$ the frequency of the driving lasers. However, radiation of such frequency escapes even from a solid-density plasma with negligible interactions, and it is of incoherent nature being of such small wavelength $\lambda_{\rm rad}=2\pi c/\omega_{\rm rad}$ that $n_e\lambda_{\rm rad}^3\ll 1$. Thus, RF losses in a laser-plasma interaction are simply measured by the amount of energy which ``disappears'' from the simulations.\footnote{In principle also low-frequency, coherent radiation which is resolved in the simulation contributes to the RF effect, thus there is some double counting of such radiation in the force since it is included both in the Lorentz and in the LL terms. However, for highly relativistic electrons with $\gamma\gg 1$ the contribution of the low-frequency part is negligible with respect to that of the dominant frequencies in the radiation spectrum.}

While a kinetic approach is most of the times necessary for a comprehensive study of laser-plasma interaction phenomena, the simplified description based on moments of Eq.(\ref{eq:kinetic}), i.e. on fluid equations, provides a suitable ground for basic models. As the motion of electrons is dominated by the superintense EM fields, one may neglect the ``random'' or thermal component of the motion and the associated pressure term, and obtain a closed set of moment equations. This is named the ``cold'' fluid approximation although the name might sound funny for such a high energy density plasma. Introducing the electron density $n_e=n_e(\rv,t)$ and fluid momentum $\pv_e=\pv_e(\rv,t)$,
\bea
n_e(\rv,t) \equiv \int f_e \dd^3p \; ,
\qquad 
\pv_e(\rv,t) \equiv n_e^{-1}\int \pv f_e \dd^3p \; ,
\eea
the cold fluid equations for electrons are 
\bea
\ddt n_e+\div(n_e\uv_e)=0 \; , 
\qquad
\frac{\dd\pv}{\dt}=(\ddt+\uv_e\cdot\grad)\pv_e=-e\left(\Ev+\frac{\uv_e}{c}\times\Bv\right) \; ,
\label{eq:coldfluid}
\eea
with $\uv_e=\pv_e/(m_e\gamma_ec)$ and $\gamma_e=(\pv_e^2+m_e^2c^2)^{1/2}$. 
Eqs.(\ref{eq:coldfluid}) are the theoretical basis for the analytic description of the laser-plasma interaction phenomena described in the following. However, in the present paper we do not enter into mathematical details.

\section{``Relativistic'' optics}

\subsection{Wave propagation and ``relativistic'' nonlinearities}

We consider a transverse EM wave ($\div\Ev=0$) propagating in an uniform plasma with electron density $n_e$. The wave equation for $\Ev$ is given by
\bea
\left(\nabla^2-\frac{1}{c^2}\ddt^2\right)\Ev
=\frac{4\pi}{c^2}\ddt\Jv
\; ,
\label{eq:waveq1} 
\eea
with the current density $\Jv=-en_e\uv_e$ (ions are assumed as an immobile, neutralizing background). For electron velocities $|\uv_e|\ll c$, we pose $\gamma_e \simeq 1$ and neglect the $\uv_e\times\Bv$ term, so that $\uv_e$ is proportional to $\Ev$. This is the basis for the linear optics of a plasma (supposed to be non-magnetized), which can be described by the refractive index ${\sf n}={\sf n}(\omega)$ with
\bea
{\sf n}^2=\varepsilon=1-\frac{\omega_p^2}{\omega^2}=1-\frac{n_e}{n_c} \; ,
\label{eq:ref}
\eea
where $\varepsilon=\varepsilon(\omega)$ is the dielectric function, $\omega_p=(4\pi e^2n_e/m_e)^{1/2}$ is the plasma frequency and $n_c=m_e\omega^2/4\pi e^2$ is named the cut-off or ``critical'' density. Wave propagation requires ${\sf n}$ to be a real number, which occurs when the wave frequency $\omega<\omega_p$ or, equivalently, the plasma density $n_e<n_c$, that defines an \emph{underdense} plasma which is transparent for the frequency $\omega$. If $n_e>n_c$ the plasma is \emph{overdense} and reflecting. For $\lambda_L= 1~\mu$m, $n_c \simeq 10^{21}$~cm$^{-3}$ which falls between the typical densities of gaseous and solid media, respectively.

When the EM wave amplitude is such that $a_0 \gtrsim 1$, nonlinear optical effects arise because of both the dependence of $\gamma_e$ on the instantaneous field and the importance of the $\uv_e\times\Bv$ term. Thus, the wave propagation depends on its amplitude and higher harmonics of the main frequency are generated. 

However, for CP there is a \emph{particular} plane wave, a monochromatic solution for which $\uv_e\times\Bv=0$ and $\gamma_e=(1+a_0^2/2)^{1/2}$ is constant in time (this solution is related to the case of the single particle orbits for CP described in sec.\ref{sec:single}). In this particular case, the electron equation of motion reduces to
\beq
\frac{\dd\pv_e}{\dt}=m_e\gamma_e\frac{\dd\uv_e}{\dt}=-e\Ev \; ,
\eeq
which is identical to the non-relativistic, linearized equation of motion but for the constant factor $\gamma_e$ that multiplies $m_e$. Thus we immediately obtain that the wave propagation can be described by the nonlinear refractive index ${\sf n}_{\rm NL}$ with
\bea
{\sf n}^2_{\rm NL}(\omega)=1-\frac{\omega_p^2}{\gamma_e\omega^2}=1-\frac{n_e}{\gamma_e n_c} \; .
\label{eq:refNL}
\eea
It should be kept in mind that, in general, a nonlinear refractive index should be used with care and that, in particular, (\ref{eq:refNL}) applies only to the idealized case of a monochromatic CP wave in a homogeneous plasma: already the extension to LP is not straightforward since $\gamma_e$ is not constant anymore. In the present context, we use (\ref{eq:refNL}) for a simple description of the phenomenon of ``relativistic'' \emph{self-focusing}. We also show, however, that applying (\ref{eq:refNL}) to the other characteristic phenomena of ``relativistic'' \emph{transparency} leads to incorrect predictions.

\subsection{Relativistic self-focusing}
\label{sec:SF}

We consider a EM beam propagating in a plasma along $x$. We assume that the beam has a standard bell-shaped profile (e.g., Gaussian), so that the intensity will be highest on the axis and decrease to zero with increasing radial distance.
$r_{\perp}$. Thus, using (\ref{eq:refNL}) as a function of the local amplitude $\av=\av(x,r_{\perp},t)$, i.e. taking $\gamma_e=(1+\braket{\av}^2/2)^{1/2}$, we obtain that ${\sf n}_{\rm NL}$ has its \emph{highest} value on the axis ($r_{\perp}=0$) and then decreases with increasing radial distance $r_{\perp}$, down to the linear value (\ref{eq:ref}). This implies that the refractive index, due to its nonlinear dependence, is modulated as in an optical fiber or dielectric waveguide, leading to a \emph{self-focusing} (SF) effect which counteracts diffraction.

\begin{figure}[t!]
\begin{center}
\includegraphics[width=0.8\textwidth]{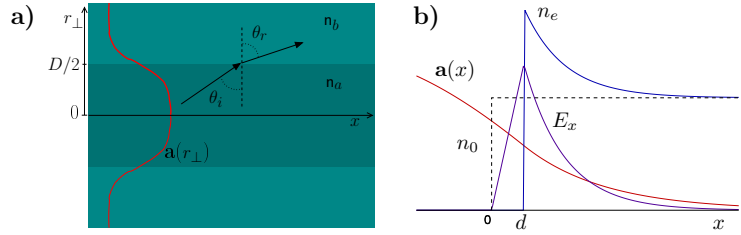}
\end{center}
\caption{a): ``optical fiber'' model of self-focusing. Since the laser beam has a radial intensity  profile $\av(r_{\perp})$ the nonlinear refractive index has higher values in the central region, causing a guiding effect. b): evanescence of the EM field $\av(x)$ in an overdense plasma ($n_0>n_c$) that fills the $x>0$ region. The electron density ($n_e$) profile is modified self-consistently by the action of the ponderomotive force which is balanced by the space-charge field $E_x$.}
\label{fig:reloptics}
\end{figure}

Fig.\ref{fig:reloptics}~a) describes a simple SF model based on a geometrical optics description. We assume a ``flat top'' radial profile so that the intensity is almost constant in the central region. Thus, the refractive index has values ${\sf n}_a={\sf n}_{\rm NL}[\av(r_{\perp}=0)]$ for $r_{\perp}<D/2$, where $D$ is the beam diameter, and ${\sf n}_b={\sf n}_{\rm NL}[\av=0]$ for $r_{\perp}>D/2$. Because of diffraction, light rays tend to diverge with a typical angle $\theta_i\simeq \arccos(\lambda/D)$. At the $r_{\perp}=D/2$ boundary, due to Snell's law the rays are bent to an angle ${\theta_r}=\arcsin\left(({\sf n}_{a}/{\sf n}_{b})\sin\theta_i\right)$, with total internal reflection occurring as $\theta_r=\pi/2$. This yields a threshold for the guiding of the beam inside the central region. In the limit of weak nonlinear effects ($|\av|\ll 1$) and small angles ($\lambda/D\ll 1$) the condition can be written as
\beq
\pi\left(\frac{D}{2}\right)^2|\av(r_{\perp}=0)|^2 \simeq \pi\lambda^2\frac{n_c}{n_e} \; .
\eeq
Note that the first term is proportional to the beam \emph{power}. Inserting numbers and recalling that $\av=e\Av/m_ec^2$, one obtains the threshold power value as $P_T \simeq 43~\mbox{\rm GW}({n_c}/{n_e})$. Thus, this rough model predicts the same scaling with density and order of magnitude as the reference value $P_T=17.5~\mbox{\rm GW}({n_c}/{n_e})$ which is obtained from a more rigorous theory \cite{sunPF87}. Notice, however, that also this latter estimate is based on some assumptions, i.e. a CP beam which is several wavelengths wide and long: it may not be applied to ultrashort, tightly focused pulses extending only over a few wavelengths. Also notice that the evolution of a laser pulse undergoing SF may be quite complex; at least, it involves the creation of a low-density channel as the electrons are pushed away from the axis due to ponderomotive forces (see sec.\ref{sec:PF}).

\subsection{Relativistic transparency}

Eq.(\ref{eq:refNL}) implies that ${\sf n}_{\rm NL}$ is real for $n_e>\gamma_e n_c$, i.e. the cut-off density is increased by a factor $\gamma_e$ with respect to the linear, non-relativistic case. The usual description is that a plasma may become transparent because of relativistic effects, and one often reads of a ``relativistically corrected'' cut-off density $\gamma_e n_c$. 

Indeed, there are two examples of ``relativistic'' transparency which are of practical importance and where taking $n_e<\gamma_e n_c$ as a criterion for wave propagation leads to erroneous predictions. The first is the case of wave incidence on a semi-infinite plasma with a step boundary. In the linear regime, one may assume the profile of the electron density to be unperturbed, so the problem is reduced to imposing boundary conditions at the plasma-vacuum interface which leads to Fresnel formulas (\cite{jackson}, par.7.3). For strong fields, however, the density profile is modified by the wave action. Taking the simplest case of normal incidence of a CP wave \cite{cattaniPRE00}, the steady ponderomotive force originating from the cycle-average of the ${\uv_e\times\Bv}$ term pushes the electrons inside the target and pile them up causing a local increase of the density in the evanescence layer, which counteracts the relativistic effect (Fig.\ref{fig:reloptics}~b). As a consequence, the threshold for wave penetration (for $n_e\gg n_c$ and $a_0\gg 1$) becomes $a_0>(\sqrt{3}/2)^3(n_e/n_c)^2$ \cite{cattaniPRE00}, which corresponds to much higher intensities than predicted by posing $\gamma_e> n_e/n_c$ i.e. $a_0>\sqrt{2}n_e/n_c$.

The second example is that of a thin foil of thickness $\ell\ll\lambda=2\pi c/\omega$, for which the relevant parameter for transparency is the areal density $n_e\ell$. The nonlinear transmission and reflection coefficients can be calculated for a normally incident CP wave by assuming a Dirac delta-like profile \cite{vshivkovPoP98}, showing the onset of transparency when
\beq
a_0>\zeta \equiv \pi \frac{n_e}{n_c}\frac{\ell}{\lambda} \; .
\label{eq:SITthin}
\eeq
Thus, for ultrathin targets such that $\ell\ll\lambda$ it is possible to have the onset of transparency even when $n_e>\gamma_e n_c$. 

It is worth noticing, however, that also these models are one-dimensional, i.e. based on plane waves. Multi-dimensional effects play an important role for any realistic laser pulse with a finite transverse profile. In particular, the ponderomotive force may reduce the electron density on axis by pushing electrons away, enhancing the penetration of the laser pulse.

\section{Interaction with a step boundary plasma}

We now focus on the interaction of a superintense laser pulse with a strongly overdense plasma ($n_e\gg n_c$) having a step-like density profile, e.g. $n_e\simeq n_0\Theta(x)$ with $\Theta(x)$ the Heaviside step function. This problem is relevant to experiments on the interaction of ultrashort pulses with solid targets. 

\subsection{Energy absorption: from Fresnel formulas to ``vacuum heating''}
\label{sec:VH}

In the linear regime, the solution for the problem of the interaction between a plane EM wave and a medium having refractive index ${\sf n}$ and a steep interface is provided by the matching relations for the wavevectors and by Fresnel formulas for the reflection and absorption coefficients, which depend on the angle of incidence and the wave polarization. Using (\ref{eq:ref}) for ${\sf n}$ one finds that inside the medium ($x>0$, for definiteness) the wave is evanescent as $\mbox{e}^{-x/\ell_s}$ with $\ell_s=c(\omega_p^2-\omega^2)^{-1/2}$ and there is total reflection of the incident energy since ${\sf n}$ is purely imaginary, which corresponds to neglecting any dissipative process. Dissipation may be provided by resistivity due to Coulomb collisions between electron and ions (Drude model), so that (\ref{eq:ref}) is modified by replacing $\omega^2\rightarrow \omega(\omega+i\nu_{\rm ei})$ where $\nu_{\rm ei}$ is the collision frequency. However, $\nu_{\rm ei}$ quickly decreases with increasing electron energy (``runaway effect'') making collisional absorption inefficient at high intensities.

  Actually, there are \emph{collisionless} mechanisms taking place in the surface region of evanescent field (the ``skin layer'') which may produce a sizable absorption (see e.g. \cite{rozmusPoP96} and references therein). 
The essence of such mechanisms is that in crossing the skin layer an electron sees the evanescent field to change in a time shorter than the oscillation period $2\pi/\omega$, so that $\braket{\vv(t)\cdot\Ev(x=x(t),t)}\neq 0$ over the electron trajectory $x(t)$. Calculating the total absorption requires a kinetic approach. However, to some extent, collisionless skin layer absorption might be included phenomenologically in the Fresnel modeling by replacing $\nu_{\rm ei}$ with an effective collision frequency. 

\begin{figure}[t!]
\begin{center}
\includegraphics[width=0.8\textwidth]{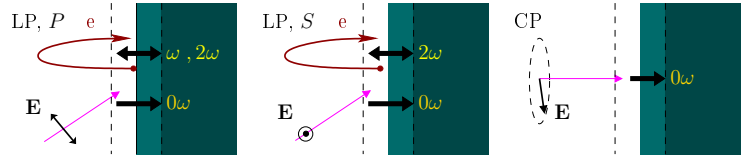}
\end{center}
\caption{Oscillatory and steady forces on an overdense plasma with steep boundary, for different polarizations. For linear polarization (LP) with $\Ev$ in the plane of incidence ($P$-polarization), both the $\Ev$ and $\vv\times\Bv$ terms in the Lorentz force can drive electron ``half-oscillations'' across the plasma-vacuum interface at a rate $\omega$ and $2\omega$, respectively. For $\Ev$ perpendicular to the plane of incidence ($S$-polarization) only the $\vv\times\Bv$ term drives the half-oscillations. For circular polarization (CP) and normal incidence, all the oscillating force components perpendicular to the surface are suppressed. In all cases, there is a steady (``$0\omega$'') force pushing the electrons and giving rise to radiation pressure action on the plasma.}
\label{fig:w2w0w}
\end{figure}

Indeed, at very high intensities absorption may be due to the generation of energetic electrons through a mechanism which violates a basic underlying assumption of the Fresnel modeling, i.e. that all electrons remain into the $x>0$ region initially occupied by the plasma. Depending on the EM wave polarization, there can be an oscillating Lorentz force component perpendicular to the surface, so that for strong enough fields an electron can be driven from the plasma surface into the vacuum region (Fig.\ref{fig:w2w0w}). After half a period of the driving force, the electron re-enters into the plasma region with a finite velocity and may cross the evanescence layer, thus escaping from the accelerating field region and being ``absorbed'' in the plasma. 
During the half-oscillation on the vacuum side, the electron acquires an energy of the order of the oscillation energy in the wave field\footnote{This estimate for the electron energy is commonly refereed to as ``ponderomotive scaling''; probably, the name originates from the questionable definition of nonlinear oscillating forces as ``ponderomotive'' (sec.\ref{sec:PF}).}, i.e. ${\cal E}_e \simeq m_ec^2\left((1+\braket{\av^2})^{1/2}-1\right)$. This is the essential description of the mechanism originally proposed by Brunel \cite{brunelPRL87} and widely referred to as ``vacuum heating'' (VH). Brunel originally considered the electric field component for $P$-polarization as the driver for electron half-oscillations across the surface, so that energetic electron bunches are generated once per laser cycle. A simple model \cite{gibbon-book} yields for the reflectivity $R$ the following implicit relation
\beq
R \simeq 1-\frac{1+\sqrt{R}}{\pi a_0}\left(\left(1+(1+\sqrt{R})^2a_0^2\sin^2\theta_i\right)^{1/2}-1\right)\frac{\sin\theta_i}{\cos\theta_i} \; ,
\eeq
with $\theta_i$ the incidence angle. In the $a_0\sin\theta_i\ll 1$ limit, 
$R \simeq 1-(4/\pi)a_0\sin^3\theta_i/\cos\theta_i$.

The magnetic component of the Lorentz force can also act as driver, so that VH may take place also for $S$-polarization and normal incidence generating electron bunches twice per laser cycle (since the magnetic force term has frequency $2\omega$).  This is also refereed to as ``$\Jv\times\Bv$'' heating, although the name comes from an earlier suggestion about the contribution of the magnetic force to absorption \cite{kruerPF85}. 
Instead, for \emph{circular} polarization and normal incidence there is no oscillating component normal to the surface\footnote{This is analogous to the absence of high-frequency longitudinal motion in a CP wave, sec.\ref{sec:planewave}.} so that electron heating may be suppressed \cite{macchiPRL05}.

\subsection{Momentum absorption and radiation pressure}
\label{sec:momentum}

\begin{figure}[t!]
\begin{center}
\includegraphics[width=0.5\textwidth]{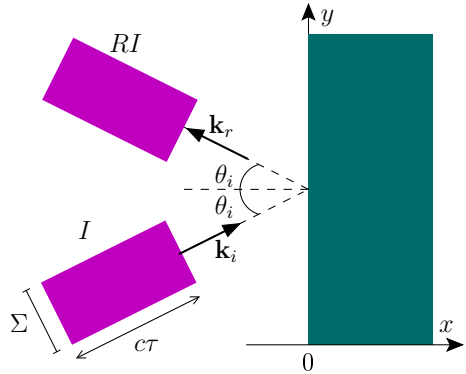}
\end{center}
\caption{Simple kinematic model to calculate the EM momentum transfer through a reflecting surface and the resulting radiation pressure. A ``box-shaped'', quasi-plane wave pulse of intensity $I$, duration $\tau$ and transverse section $\Sigma$ impinges at an angle $\theta_i$ on the surface. If the latter is at rest, the reflected pulse is in the specular direction, has the same duration and section as the incident pulse, and an intensity $RI$ where $R$ is the reflectivity of the surface.}
\label{fig:oblref}
\end{figure}

In addition to energy, EM field contain traslational momentum, its density being $\gv=\Ev\times\Bv/4\pi c$. Thus, an idealized quasi-plane-wave ``square'' pulse of duration $\tau$ and transverse area $\Sigma$ (Fig.\ref{fig:oblref}) contains a total momentum $\pv_i=\gv\Sigma c\tau=(I/c^2)(\Sigma c\tau)\nv$ where $I=(c/4\pi)|\Ev\times\Bv|$ is the intensity and $\nv$ the direction of propagation. Under reflection from the surface of a medium with reflectivity $R$, momentum is transferred to the medium giving rise to a net force perpendicular to the surface, i.e. to radiation pressure. By simple kinematic relations, the pressure on the surface can be obtained as 
\beq
P_{\perp}=(1+R)\frac{I}{c}\cos^2\theta_i \; ,
\eeq 
where we took $\nv=(\cos\theta_i,\sin\theta_i)$ and the surface at $x=0$. The maximum pressure of $2I/c$ is obtained for a perfect mirror ($R=1$) at normal incidence ($\theta_i=0$). The above relations are of classical nature, however one may also obtain the radiation pressure kinematically by describing the incident pulse as a bunch of $N$ photons each of energy $\hbar\omega$ and momentum $(\hbar\omega/c)\nv$ of which a fraction $R$ is elastically reflected at the surface. The classical expression is recovered by the equation for the pulse/bunch energy $I\Sigma c\tau=N\hbar\omega$. 

Going back to the classical description, one can also obtain the total pressure from the knowledge of the EM fields by integrating the total force per unit volume over the whole plasma,
\beq
P_{\perp}=\int_0^{+\infty}\left(\rho\Ev+\frac{\Jv}{c}\times\Bv\right)\cdot\xv\dd x \; ,
\label{eq:Pperp}
\eeq
where $\rho$ is the charge density. To test a simple case, we may assume normal incidence $(\theta_i=0)$ so that $\Ev\cdot\xv=0$, and calculate the fields inside the plasma in the linear limit by using Fresnel formulas with ${\sf n}$ given by (\ref{eq:ref}) so that $R=1$. In this case, besides recovering easily the result $P_{\perp}=2I/c$ one observes that the integrand of (\ref{eq:Pperp}) is the non-relativistic ponderomotive force (\ref{eq:PF}) multiplied by $n_e$. 
In practice the local ponderomotive force is on the electrons only (the $\vv\times\Bv$ on ions is smaller by a factor $\sim m_e/m_i \sim 10^{-3}$), 
but as soon as the force pushes the electrons in the region of evanescent fields, a charge depletion layer is created at the surface with an electrostatic field which back-hold electrons and exerts a force on ions in the inward direction. This situation is evidence in Fig.\ref{fig:reloptics}~b) which shows the charge separation layer ($0<x<d$) and the corresponding electrostatic field $E_x$.
If the electrons are in equilibrium, the ponderomotive force is exactly balanced locally by the electrostatic one, so in turn the ions feel an electrostatic pressure which equals the radiation pressure value. In the absence of counteracting forces, the electrostatic field will accelerate ions, so that ultimately the EM momentum is transferred to the whole medium. Radiation pressure of superintense lasers is currently investigated as a driving mechanism for laser-plasma accelerators of ions \cite{macchiRMP13}: related concepts are investigated in sec.\ref{sec:mirrors}.

\subsection{Absorption of tangential momentum}
\label{eq:transversemomentum}

By applying the same kinematics leading to Eq.(\ref{eq:Pperp}), we also obtain that for a medium with partial reflectivity ($R<1$) there is absorption of EM momentum also in the \emph{parallel} direction, i.e. along the surface, yielding a \emph{tangential} pressure.  
\bea
P_{\parallel}=(1-R)\frac{I}{c}\sin\theta_i\cos\theta_i \; . 
\label{eq:Ppar}
\eea
We thus expect that (referring to the two-dimensional, plane wave geometry of Fig.\ref{fig:oblref}) the ponderomotive force has a tangential ($y$) component $F_{py}$, which can drive a surface current $j_y$ of electrons. Such surface current has been often observed in simulations since early studies of absorption at oblique incidence \cite{brunelPF88} but, to our knowledge, no simple model was presented until recently; below we resume the basic findings of our model \cite{macchiXXX19} which were partly anticipated in Ref.\cite{grassiPRE17}.

If the plasma is homogeneous along $y$, the current $j_y$ produces no charge separation and thus no electrostatic field. Indeed, $j_y$ generates a magnetic field $B_z$ which, while growing in time, induces an electric field $E_y$ which counteracts the ponderomotive action. However, the evanescence lengths of $F_{py}$ and $E_y$ are different, so that the ponderomotive and electric forces cannot balance locally and a double layer of current is generated, which leads to a $B_z$ localized in the skin layer. For an incident EM wave with flat-top profile, i.e. having constant intensity $I=I_0$ for $0 \leq t < \tau_L$, both $j_y$ and $B_z$ are found to grow linearly in time until $t=\tau_L$ with the maximum value of $B_z$ at the time $t$ being
\beq
B_z^{\rm (max)} \simeq \frac{\pi}{6}\frac{t}{\tau_L}(1-R)\sin(2\theta_i)a_0B_L \; ,
\eeq
where $a_0=I_0/m_en_cc^3$ and $B_L$ are the dimensionless and magnetic field amplitudes, respectively, of the incident wave. Intense laser pulses ($a_0\gg 1$) can yield high absorption and low reflectivities down to $R \simeq 0.5$, so that the amplitude of the slowly-varying field $B_z$ may approach that of the laser field $B_L$, i.e. $\simeq 10^9$~Gauss for $a_0 \sim 10$.

\section{Moving mirrors}
\label{sec:mirrors}

The picture of ``vacuum heating'' presented in sec.\ref{sec:VH}, in which electrons are periodically dragged out of and back into the plasma, is oversimplified. In reality the oscillating components of the Lorentz force drive a collective oscillation of the electron density profile (with the high energy electron bunches being related to the partial ``breaking'' of such oscillations). We may thus assume that the  $n_e=n_c$ surface oscillates back and forth under the action of the Lorentz force. Thus, the incident laser pulse is reflected from a surface whose position oscillates either at the same frequency of the laser, or twice that value depending on the incidence angle and polarization. 
If we consider instead the action of the time-averaged force, i.e. of radiation pressure, the $n_e=n_c$ surface is pushed inwards, so we have reflection from a surface moving along the propagation direction. The relativistic \emph{moving mirror} model is able to explain (at least qualitatively) basic features of both the above mentioned scenarios, which are relevant to important applications of superintense interaction with overdense plasmas (e.g. solid targets). It is thus worth to review here some basic relations of reflection from a moving mirror.

\subsection{Reflection from a moving mirror}

\begin{figure}[t!]
\begin{center}
\includegraphics[width=0.8\textwidth]{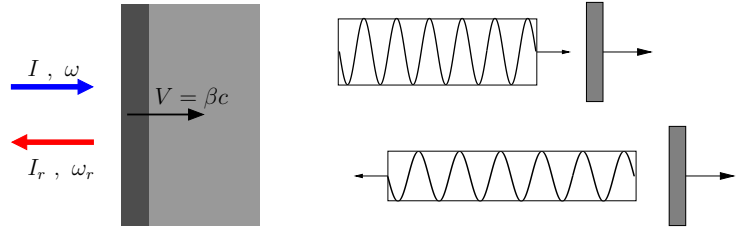}
\end{center}
\caption{a) EM wave of intensity $I$ and frequency $\omega$ impinging on a moving mirror. When the mirror velocity $\Vv$ is in the propagation direction as in the picture, the wave frequency is red-shifted and a reflected pulse has longer duration and lower energy than the incident pulse. Conversely, blue-shift and energy increase occur for a counter-propagating mirror.}
\label{fig:movingmirror}
\end{figure}

For brevity and simplicity we consider normal incidence only and we assume a ``perfect'' mirror whose reflectivity $R=1$ \emph{in its rest frame}. Let the mirror move with velocity $\Vv=V\xv$ and an EM plane wave of frequency $\omega$, field amplitude $E_i$ and intensity $I=(c/4\pi)E_i^2$ be incident from the $x<X_m$ side, where $X_m$ is the mirror position (Fig.\ref{fig:movingmirror}). For the moment we assume $V$ to be constant, hence $X_m=Vt$.

The laws of reflection are known in the rest frame of the mirror ($L'$): the EM wave is reflected with inversion of both the wavevector and the electric field and no change of frequency. Thus we can obtain the frequency $\omega_r$ and the amplitude $E_r$ of the reflected wave in the lab frame ($L$) by a first Lorentz transformation of the incident wave from $L$ to $L'$, and then by a second transformation of the reflected wave from $L'$ to $L$. The result is 
\beq
\frac{\omega_r}{\omega}=-\frac{E_r}{E_i}=\frac{1-\beta}{1+\beta} \; , 
\label{eq:movingmirror1}
\eeq
where $\beta=V/c$. Thus, if $V>0$, i.e. if the EM wave propagates in the same direction as the mirror velocity, the frequency is ``red-shifted'' towards lower values and the amplitude is also lower than for the incident pulse. If $V<0$, i.e. if the wave is counterpropagating with respect to the mirror, ``blue-shift'' and amplitude increase occur. In the highly relativistic limit ($\beta\rightarrow 1$) notice that $({1-\beta})/({1+\beta}) \simeq (2\gamma)^{-2}$.

The above relations might also be found by noticing that, for normal incidence (and thus the electric field parallel to the mirror surface) the boundary condition $\Ev'(x'=X'_m)$ for a perfect mirror at rest in $L'$  corresponds to $\Av(x=X_m)=0$ in $L$ for arbitrary motion $X_m=X_m(t)$, as can be easily demonstrated via a Lorentz transformation and the relations between $\Av$, $\Ev$ and $\Bv$. Thus, by posing
\beq
\left[A_i\mbox{e}^{ikx-i\omega t}+A_r\mbox{e}^{-ik_rx-i\omega_rt}\right]_{x=Vt}=0 \; ,
\label{eq:movingmirror2}
\eeq
where $k=\omega/c$ and $k_r=\omega_r/c$, Eqs.(\ref{eq:movingmirror1}) are obtained again. 

If we consider an incident pulse of long but finite duration $\tau$, such as the ``square'' packet in Fig.\ref{fig:movingmirror}, the number of oscillations inside the pulse is a Lorentz invariant. Thus, the duration of the reflected pulse is $\tau_r=\tau({1+\beta})/({1-\beta})$, i.e. $\tau_r>\tau$ if $V>0$ and $\tau_r<\tau$ if $V<0$. Since the intensity of the reflected field is $I_r=I({1-\beta})^2/({1+\beta})^2$, we find that $I_r\tau_r<I\tau$ for $V>0$, i.e. the incident pulse loses energy to the mirror, while the opposite occurs for $V<0$. A counterpropagating mirror may thus be used to both compress in time and amplify an incident pulse: an intriguing laser-plasma based scheme of such kind has been proposed as a way to reach unprecedentedly high intensities \cite{bulanovPRL03}.

\subsection{High harmonics from an oscillating mirror}

\begin{figure}[t!]
\begin{center}
\includegraphics[width=0.8\textwidth]{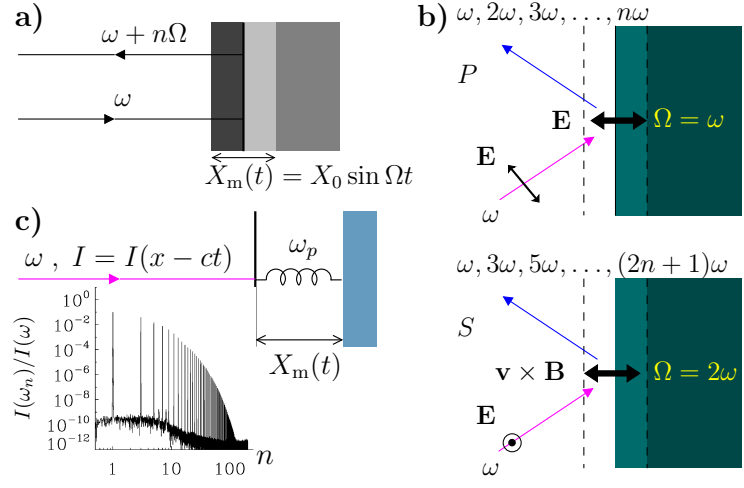}
\end{center}
\caption{Oscillating mirrors and harmonic generation. a): frequency mixing in the reflected wave. b): driving of a plasma surface at different frequencies depending on the polarization and incidence angle, leading to the generation of different order of harmonics. c): a toy model for a laser-driven oscillating mirror. The inset shows a spectrum of the reflected pulse obtained with such a model ($n$ is the harmonic order).}
\label{fig:HH}
\end{figure}

Now suppose the perfect mirror performs an oscillatory motion, $X_m=X_0\sin\Omega t$. To find the reflected field we can use again the condition $A(x=X_m,t)=0$ and thus write, e.g., 
\beq
0=[A_i(x,t)+A_r(x,t)]_{x=X_0\sin\Omega t}=[A_i\cos(kx-\omega t)+A_r(x,t)]_{x=X_0\sin\Omega t} \; ,
\eeq
from which we obtain, using some math, that the temporal dependence of the reflected pulse is
\beq
 A_r(t) \sim \sin\left(\omega t+\frac{2\omega}{c}X_0\sin\Omega t\right)
\sim \sum_{n=0}^{\infty}J_n\left(\frac{2\omega X_0}{c}\right)\sin(\omega+n\Omega)t \; ,
\eeq
where the $J_n$'s are Bessel functions. Thus, the reflected wave contains a mixing of $\omega$, the frequency of the incident wave, with integer harmonics of the mirror frequency, $\Omega$. 

An intense laser pulse of frequency $\omega$ drives oscillations of the surface of an overdense plasma at frequency $\omega$ or $2\omega$ depending on the angle of incidence and the polarization (Fig.\ref{fig:HH}~b). The moving mirror model thus predicts that a $P$-polarized pulse will generate $P$-polarized harmonics at all integer frequencies of the driving pulse ($\omega,~2\omega,~3\omega, \ldots$) while a $S$-polarized pulse will generate only odd frequencies $(2n+1)\omega$. Of course, since the mirror is driven by the same laser pulse it reflects, any estimate of the intensity of such harmonics must be based on some self-consistent modeling for dynamics of the moving mirror. A toy model might be formulated by assuming that the mirror is bound by a spring of frequency $\omega_p$ (Fig.\ref{fig:HH}~c), which roughly accounts for the resonant plasma response, and by inserting a friction term to phenomenologically account for finite absorption. For a mirror driven by a linearly polarized, ``flat-top'' (constant intensity $I$) pulse at normal incidence, the equation of motion is
\beq
\frac{\dd}{\dt}(\gamma_m\beta_m)=\frac{2I}{\sigma c^2}\left(1+2\cos(2\omega t_r)\right)\frac{1-\beta_m}{1+\beta_m}-\omega_p^2X_m -\nu_m\beta_m c \; ,
\label{eq:MM}
\eeq
where ${\dd X_m}/{\dt}=\beta_m c$ and $t_r=t-X_m/c$. In (\ref{eq:MM}) $\sigma$ is the mass per unit area of the mirror, so that when referring to an oscillating plasma surface we might roughly estimate $\sigma \simeq m_en_e\ell_s$ with $\ell_s$ the evanescence length (ions are assumed to be at rest).
Eq.(\ref{eq:MM}) may be easily solved numerically to obtain the maximum velocity of the mirror $\beta_{\rm max}c$, which according to (\ref{eq:movingmirror1}) should be related to the spectral cut-off frequency $\omega_{\rm co} \simeq 4\omega\gamma^2_{\rm max}$ when $\beta_{\rm max}\rightarrow 1$. Thus, if $\gamma \sim (1+a_0^2)^{1/2}$ one expects to generate harmonics up to orders $\sim 10^2$ with state-of-the-art lasers. 
One can also obtain, via (\ref{eq:movingmirror2}), the temporal profile of the reflected pulse. The latter usually appears as a train of ultrashort spikes, which can be qualitatively understood as a coherent modulation of the incident pulse waveform by the moving mirror: each semicycle is alternatively stretched or compressed depending on the sign of $\beta_m(t)$. A quantitative description of high harmonic generation needs a more realistic modeling and simulations of the laser-plasma dynamics, of course (see \cite{teubnerRMP09,thauryJPB10} for reviews).

\subsection{Light sail acceleration}
\label{sec:LS}

Now assume a thin plane mirror of mass density $\rho_m$ and thickness $\ell$, and a plane wave pulse $I=I(t)$ at normal incidence and with circular polarization so that there are no oscillating components. The mirror is thus accelerated by radiation pressure according to the equation of motion
\beq
\frac{\dd}{\dt}(\gamma_m\beta_m)=\frac{2I(t_r)}{\rho\ell c^2}R(\omega')\frac{1-\beta_m}{1+\beta_m} \; ,
\label{eq:LS}
\eeq
which we name the \emph{light sail} (LS) equation.
As we consider the acceleration of the foil as a whole\footnote{Note that $\rho_m\ell$ in Eq.(\ref{eq:LS}) is formally equivalent to $\sigma$ in Eq.(\ref{eq:MM}), but here in (\ref{eq:LS}) $\rho_m\ell$ refers to the total mass of the mirror, i.e. including the ions.}, with respect to Eq.(\ref{eq:MM}) there are no elastic and friction terms. Instead, we include a finite reflectivity $R<1$ to account for partial transmission through the foil. Notice that in general $R$ depends on the incident pulse frequency and it is defined for a mirror at rest, thus it is a function of the frequency in the moving frame $\omega'=\omega(1-\beta_m)^{1/2}(1+\beta_m)^{-1/2}$ and, for a thin ($\ell\ll\lambda$) plasma mirror it is proportional to $\rho\ell$. 
At intensities high enough for relativistic transparency effects to be important, $R$ quickly drops from unity as the threshold in Eq.(\ref{eq:SITthin}) is exceeded, so that $a_0\simeq\zeta$ is an optimal compromise  between reducing the areal mass and increasing reflectivity at fixed thrust in order to maximize the sail acceleration. In the following we assume for simplicity $R=1$ although an analytic solution of Eq.(\ref{eq:LS}) may be found also for a partially transparent ``delta-like'' foil \cite{macchiNJP10}.

From Eq.(\ref{eq:LS}) the final $\gamma$-factor is obtained as 
\beq
\gamma_m(t=\infty)-1= \frac{{\cal F}^2}{2({\cal F}+1)} \; ,  \qquad 
{\cal F}=\frac{2}{\rho\ell}\int_0^{\infty}I(t')\dt' 
\; ,
\eeq
where ${\cal F}$ can be estimated as a function of the average intensity $I$ and pulse duration $\tau$,
\beq
{\cal F}=\frac{2{I}\tau}{\rho\ell}=\frac{Z}{A}\frac{m_e}{m_p}\frac{a_0^2}{\zeta}\omega\tau \; .
\eeq
We thus see that present-day femtosecond lasers having $\tau \sim 10(2\pi/\omega)$ and $a_0 \sim 10$ are in principle able to accelerate ultrathin targets up to $\gamma_m-1 \gtrsim 0.1$, which corresponds to an energy per nucleon exceeding 100~MeV, while future lasers yielding $a_0 \sim 10^2$ could drive relativistic GeV nuclei. In addition, LS acceleration becomes more efficient with increasing speed, the mechanical efficiency $\eta_{\rm mec}$ (ratio of sail energy ${\cal E}_{\rm LS}$ over driver pulse energy $I\tau$, all defined per unit surface) being
\beq
\eta_{\rm mec} \equiv \frac{{\cal E}_{\rm LS}}{I\tau}=\frac{2\beta_m}{1+\beta_m} \; .
\eeq
This relation can be obtained from Eq.(\ref{eq:LS}), but also from a simple quantum picture taking the pulse as a bunch of ${\cal N}$ photons (per unit surface) whose energy drops from $\hbar\omega$ to $\hbar\omega_r$ due to reflection from the sail. Thus, since ${\cal N}=I\tau/\hbar\omega$, 
\beq
{\cal E}_{\rm LS}={\cal N}\hbar(\omega-\omega_r)={\cal N}\hbar\omega\frac{2\beta_m}{1+\beta_m}=\eta_{\rm mec} I\tau \; .
\eeq
The efficiency of LS acceleration is what makes it attractive for interstellar propulsion of probes from Earth \cite{meraliS16} as well for laser-driven ion accelerators \cite{macchiRMP13}. For this latter application, additional features as monoenergetic spectrum and ultrashort duration (since ideally all ions in the sail propagate at the same velocity) make the LS appear as a ``dream bunch'' of energetic ions. Issues include the slow energy gain, since Eq.(\ref{eq:LS}) shows that the force on the sail decreases with increasing $\beta_m$ so that reaching the highest possible energy requires stability over long distances. The modeling in a realistic geometry brings both good news (LS might be faster and more efficient in 3D than in 1D \cite{bulanovPRL10,sgattoniAPL14}, which is uncommon) and bad news (the sail might be prone to Rayleigh-Taylor-type instabilities \cite{sgattoniPRE15,eliassonNJP15}, see Section~\ref{sec:instabilities}).

\section{Instabilities}
\label{sec:instabilities}

\emph{Instability} is maybe the word which is more frequently associated to \emph{plasma}, the obvious reason being that the main obstacle to achieving controlled fusion is that a plasma tends to become unstable in several ways, quickly destroying the desired configuration. The basic laser-plasma interaction processes we reviewed so far (as well as other we did not include) may also lead to, or be affected by instabilities. For example, a laser pulse greatly exceeding the power threshold for relativistic self-focusing may break up in multiple filaments, especially if its intensity distribution is not smooth. As another example, the high-energy electrons produced by laser-plasma interactions typically lead to an anisotropical distribution function which is unstable against electromagnetic perturbations (Weibel instability): the growth of the latter act to deviate particle trajectories in order to create a more isotropic distribution. In the context of laser-plasma interactions one also encounters nonlinear processes where a strong ``pump'' mode having frequency $\omega_0$ and wavevector ${\bf k}_0$, such as e.g. an intense laser pulse propagating in the plasma or an high amplitude plasma wave, excites two (or more) ``daughter'' plasma modes whose frequencies and wavevectors are related by the phase matching relations $\omega_0=\omega_1+\omega_2$ and ${\bf k}_0={\bf k}_1+{\bf k}_2$. These processes are referred to as \emph{parametric instabilities} since the daughter modes may also grow at high amplitude at a rate typically proportional to the amplitude of the pump mode. An example is Raman backscattering with corresponds to a laser wave exciting a plasma wave and an EM wave in the backward direction, which can lead to strong reflection from a low density plasma.

Covering all the possible instabilities in the laser-plasma scenario is much beyond the limits and scope of the present paper, thus we just give some further detail on instabilities affecting the dynamics of the moving mirror dynamics outlined in Section~\ref{sec:mirrors}. The plasma surface oscillating under the action of the Lorentz force has been found in simulations to develop ripples which also oscillate at half the driving frequency \cite{macchiPRL01}. This is due to a parametric instability in which the driven surface oscillation decays into two surface waves, similarly to the phenomenon of Faraday ripples (or waves)\footnote{\url{https://en.wikipedia.org/wiki/Faraday_wave}} originating on the surface of a fluid subject to vertical vibrations. In the context of laser-plasma interaction the effect was studied in relation to the onset of surface rippling in experiments on high harmonic generation, where the harmonic emission was observed to turn from collimated to diffuse over a certain intensity threshold.  

\begin{figure}[t!]
\begin{center}
\includegraphics[width=0.3\textwidth]{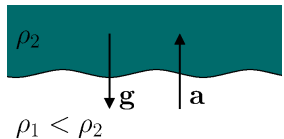}
\end{center}
\caption{Rayleigh-Taylor instability: an interface between two fluids of different mass density becomes corrugated in the presence of a gravity field anti-parallel to the density gradient or, equivalently, an acceleration parallel to the density gradient.
}
\label{fig:RT}
\end{figure}

When the plasma surface is steadily accelerated by radiation pressure as in the light sail concept (Section~\ref{sec:LS}), rippling may occur because of an instability of the Rayleigh-Taylor (RT) type. The simplest example of RT instability (RTI) is that of an heavy fluid of density $\rho_2$ placed above a lighter one of density $\rho_1<\rho_2$ in a gravity field ${\bf g}$ (Fig.\ref{fig:RT}): a small perturbation at the surface lowers the energy of the system and thus grows up exponentially ($\sim \mbox{e}^{\gamma_{\rm RT}t}$) in a first stage, favoring the mixing of the two fluids. The equivalence principle tells us that the same effect is produced in the presence of an acceleration field ${\bf a}$ directed from the light fluid to the heavier one: this is the instability form which strongly affects the compression of fuel pellet in Inertial Confinement Fusion \cite{atzeni-book}.

For a sinusoidal perturbation of wavevector $k_{\rm RT}$, the RTI growth rate is given by (see e.g. \cite{chandrasekhar-hydrobook})
\begin{equation}
\gamma_{\rm RT}=\left(ak_{\rm RT}\frac{\rho_2-\rho_1}{\rho_2+\rho_1}\right)^{1/2} \; ,
\end{equation}
where $a=|{\bf g}|$ in the case of the gravitational RTI.
The case of a plasma surface accelerated by radiation pressure can be viewed as a massless fluid of photons pushing a heavy material fluid, and it is thus unstable with a rate $\gamma_{\rm RT}=(ak_{\rm RT})^{1/2}$. RTI also occurs for a thin interface layer separating two fluids of different pressures, which matches closely the LS scenario where the target is placed between the photon fluid and vacuum. The growth rate of such RTI, for non-relativistic dynamics, has the same form as the preceding formula with $a=(2I/\rho\ell c)$ \cite{ottPRL72}. Analytical models accounting for relativistic motion and other effects can be found, e.g., in Refs.\cite{pegoraroPRL07,khudikPoP14}.  These works left open the question why the surface rippling often observed in simulations occurs predominantly for a wavevector $k_{\rm RT}\simeq 2\pi/\lambda$, i.e. with a periodicity close to the laser wavelength. In Refs.\cite{sgattoniPRE15,eliassonNJP15} it has been suggested that the rippling of the surface self-modulates the radiation pressure, so that depending on the laser polarization the accelerating force may become stronger in the valleys of the ripples and boost their growth. The effect is strongest for a sinusoidal rippling at the laser wavelength because of a resonant coupling with surface plasma waves.  

\section{Angular momentum absorption and magnetic field generation}

The fact that an EM wave carries energy and momentum becomes very eye-catching for superintense laser pulses which, as we saw in the preceding section, can heat matter to extremely high temperatures and accelerate a quite macroscopic object to velocities approaching the speed of light. An EM wave with CP also carries angular momentum which, when absorbed by a sample of matter, may cause its rotation. For a CP laser beam of frequency $\omega$, propagating along $x$ and having a radial intensity profile $I(r)$, the density of angular momentum along the $x$-direction is
\beq
{\cal L}_x=(\rv\times\gv)_x=-\frac{r}{2c\omega}\ddr I(r) \; ,
\eeq 
where $\gv$ is the density of traslational momentum (sec.\ref{sec:momentum}). Notice that for a standard bell-shaped profile ${\cal L}_x$ peaks at the edge of the beam. The total angular momentum $L_x$ is proportional to the power $P$ of the beam,
\bea
L_x=\int_0^{\infty}{\cal L}_x(r)2\pi r\dd r
=\frac{1}{c\omega}\int_0^{\infty}I(r)2\pi r\dd r=\frac{P}{c\omega} \; .
\eea
We have seen in sec.\ref{sec:mirrors} than in the reflection from a perfect mirror an EM wave delivers twice of its traslational momentum, and that if the mirror moves at relativistic velocities most of the EM wave energy is converted into mechanical energy of the mirror. However, it can be shown that \emph{no} angular momentum is transferred to the mirror. The reasoning is very simple by taking a quantum point of view: the value of the ``spin'' angular momentum of a photon is $\hbar$, independently of the frequency, and in the reflection the spin is not reversed while the number of photons is conserved for a perfect mirror, so there is no net absorption of angular momentum. 

In general, absorption of EM angular momentum requires a dissipative mechanism which ``destroys'' part of the incident photons. At moderate intensities such mechanism is provided by collisions \cite{hainesPRL01}. At extremely high intensities, strong losses by incoherent emission of radiation imply the absorption of many laser photons for each high frequency photon emitted, hence the transfer of angular momentum might become very efficient in a regime dominated by radiation friction effects \cite{liseykinaNJP16}. 

The angular momentum of a laser beam is directly absorbed by the electrons, and the associated torque drives an azimuthal electron current. In turn, this current generates an axial magnetic field: this is known as the \emph{inverse Faraday effect} (IFE) even if this is somewhat a misnomer. Even with a steady absorption, the axial field cannot grow indefinitely since it is accompanied by the induction of a solenoidal electric field that counteracts the electron rotation and exerts a torque on ions, which ultimately absorb most of the angular momentum. The mechanism is thus similar to that leading to the absorption of transverse momentum (sec.\ref{eq:transversemomentum}).
The scaling of the peak magnetic field on axis $B_{\rm ax}$ with laser and plasma parameters is found to be \cite{hainesPRL01,liseykinaNJP16}
\beq
B_{\rm ax} \sim \eta \frac{n_c}{n_e}\frac{c\tau\lambda^2}{D^2L} B_0a_0^2 \; ,
\eeq 
where $\eta$ is the absorbed fraction of the laser energy, $L$ is the length over which absorption occurs, $B_0=m_ec\omega/e$ and other parameters are as previously defined. Notice that $B_0a_0=B_L$, the magnetic field amplitude of the laser pulse. Simulations with radiation friction included \cite{liseykinaNJP16} of the interaction of superintense pulses with overdense plasmas have shown strong radiation losses with $\eta$ up to 25\% and a scaling $\eta \sim a_0^3$, so that $B_{\rm ax} \sim a_0^4$. In the simulated conditions, which might be accessible with next-generation lasers, the generation via IFE of magnetic fields of several $10^9$~Gauss is observed, providing in the meantime a demonstration of a macroscopic effect of radiation friction.


\end{document}